\documentclass[twocolumn,aps,pra,showpacs,floatfix]{revtex4}
\usepackage{graphicx}
\usepackage{epsfig}
\usepackage{amsmath}
\usepackage{psfig}
\usepackage{amssymb}
\def\one{\hbox{$\bigcirc$\hspace{-0.3cm}{\em 1}}\hspace{0.25cm}}
\def\two{\hbox{$\bigcirc$\hspace{-0.3cm}{\em 2}}\hspace{0.25cm}}
\def\three{\hbox{$\bigcirc$\hspace{-0.3cm}{\em 3}}\hspace{0.25cm}}

\newcommand{\egy}[1]{\begin{equation}#1\end{equation}}

\begin{document}
\title{Entanglement in the XX spin chain with an energy current}

\author{V. Eisler$^{1,\footnote{eisler@general.elte.hu}}$
and Z. Zimbor\'as$^{2,\footnote{cimbi@rmki.kfki.hu}}$}

\affiliation{${}^1$Institute for Theoretical Physics,
E\"otv\"os University, 1117 Budapest, P\'azm\'any s\'et\'any 1/a, Hungary\\
${}^2$Research Institute for Particle and Nuclear
  Physics, P.O. Box 49, H-1525 Budapest, Hungary}

\date{\today}
\begin{abstract}
We consider the ground state of the XX chain that is constrained to carry a
current of energy. The von Neumann entropy of a block of $L$ neighboring
spins, describing entanglement of the block with the rest
of the chain, is computed.
Recent calculations have revealed that the entropy in the
XX model diverges logarithmically with the size of the subsystem.
We show that the presence of the energy current increases the
prefactor of the logarithmic growth. This result indicates that the emergence
of the energy current gives rise to an increase of entanglement.

\end{abstract}


\maketitle

\section{Introduction}

Recently entanglement properties of various quantum systems
have been the focus of numerous studies. Entanglement
plays an essential role in several many-body quantum
phenomena, such as superconductivity \cite{supcond} and
quantum phase transitions \cite{Sachdev}. It is also regarded as
an important resource in quantum computation and information
processing \cite{qcomp}. Quantum spin chains offer an excellent 
theoretical framework for investigating entanglement properties,
since several simple models can be solved analytically,
and there also exist efficient numerical techniques.
This motivated us to work with spin chains in order to
investigate the effect of energy current on entanglement.
\par
There are two widely used method of characterising
entanglement in spin chains. The first of these
describes the entanglement between two spins in the chain
with the quantity called concurrence \cite{conc1,conc2}. The other
one measures entanglement of a block of spins with
the rest of the chain with the von Neumann entropy,
when the chain is in its ground state
\cite{Kitaev,Korepin,belga,Keating,Popkov}. This latter method
is especially useful when one tries to understand the role
of entanglement in quantum phase transitions. These transitions
manifest themselves in the appearance of gapless excitations,
and are accompanied by a qualitative change in the correlations.
In view of the connection between entanglement and
quantum correlations, the motivation to characterize a
critical system in terms of entanglement properties
naturally emerges.
\par
Vidal \emph{et al.} \cite{Kitaev}
calculated the von Neumann entropy for a wide range of one-dimensional 
spin models and found that for critical (gapless) ground states
the entropy of a block of spins diverges logarithmically
with the size of the block, while for noncritical chains
it converges to a finite value. The prefactor of the
logarithm was argued to be one-third of the central charge
of the underlying conformal field theory.
These results were supported by analytical calculations
for the XX chain in \cite{Korepin} and for more general
Hamiltonians in \cite{Keating}.
\par
Spin chains are simple enough models to investigate
also the nonequilibrium effects on entanglement.
One can find states that are characterised by the presence of
currents of some physical quantities such as energy or
magnetization \cite{tiwc,xxwc}. An important effect
of these currents is the rather drastic change in
correlations. Therefore introduction of a current
can be regarded as a quantum phase transition to a 
current-carrying phase. Consequently, it is interesting
to find the entanglement properties of these current-carrying states.
\par
In this paper we study an XX spin chain constrained to
carry an energy current. We calculate the von Neumann 
entropy of a subsystem of $L$ contiguous spins.
The presence of the energy current maintains the logarithmic
asymptotics of the entropy; however, the prefactor of
the logarithm is increased from 1/3 to 2/3, indicating a 
higher level of entanglement in the current-carrying states.
We also show that at a special value of the current, where the 
symmetry of the state is enhanced, the asymptotics of the entropy
is the same as in the XX chain without current. In the
vicinity of these transition points the entropy is shown to display
a special type of finite-size scaling.
\section{XX chain with energy current}
\par
The XX model is defined through the following Hamiltonian:
\egy{H^{XX}=-\sum_{l=1}^N{(s_l^x s_{l+1}^x +
s_l^y s_{l+1}^y)}- h \sum_{l=1}^N{s_l^z},}
where $s^\alpha_l (\alpha=x,y,z)$ are the Pauli spin matrices at
sites $l=1,2,\dots,N$ of a periodic chain and $h$ is the magnetic field.
Our aim is to constrain the spin chain to carry a prescribed amount
of energy current; therefore we use the technique introduced in
\cite{xxwc}. Since the local energy satisfies a continuity equation
with the local energy current, one can calculate the operator of the
total energy current:
\egy{
\begin{split}
J^E= & \sum_{l=1}^N [ s_l^z(s_{l-1}^y 
s_{l+1}^x-s_{l-1}^x s_{l+1}^y) \\ & + h (s_l^x s_{l+1}^y 
- s_l^y s_{l+1}^x) ]
\end{split}}
In order to find the lowest-energy state among the states carrying
a given current, one has to introduce a Lagrange multiplier $\lambda$,
and diagonalize the following modified Hamiltonian:
\egy{H^E=H^{XX}-\lambda J^E.}
The ground state of $H^E$ can be considered as a current-carrying
steady state of $H^{XX}$ at zero temperature.
\par
\begin{figure}[thb]
\includegraphics[width=0.7\columnwidth,angle=270]{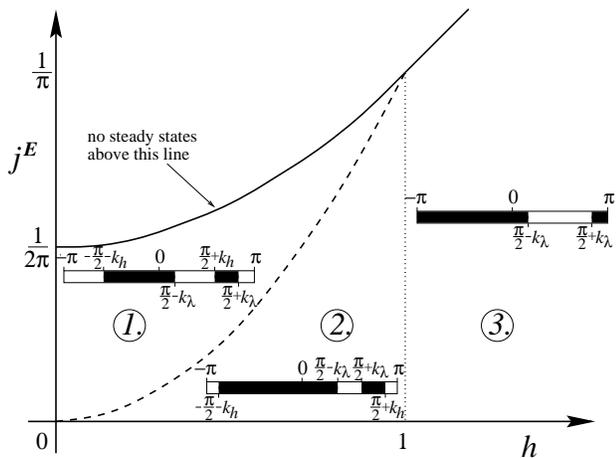}
\caption{Phase diagram of the XX model with energy current
        in the $h-j^E$ plane, where $h$ is the transverse field while
        $j^E$ is the density of the flux of energy. The occupied
        fermionic modes are represented by the black parts of the
	rectangles \cite{xxwc}.}
\label{fig:phd}
\end{figure}

Since $\left[H^{XX},J^E\right]=0$, one can diagonalize $H^E$ using the 
same methods which diagonalize $H^{XX}$ \cite{LSM}, and the model
can be transformed into a set of free fermions with the following
spectrum:
\egy{\Lambda_k=(-\cos k -h)(1-\lambda \sin k).}
The ground state can be constructed by occupying all the modes
with negative energy, and it remains the same as that of $H^{XX}$
for $\lambda\le 1$. If the driving field $\lambda$ exceeds this
critical value the energy current starts to flow, and the
Fermi sea of the occupied modes splits into two parts.
In order to illustrate the occupied regions
it is useful to introduce the characteristic wavelengths
$k_h=\arcsin(h)$ and $k_\lambda=\arccos(\lambda^{-1})$.
The ground state can be analyzed as a function of $h$ and
the expectation value of the current density
$j^E=\langle J^E/N \rangle$, and the phase diagram shown on
Fig.\ref{fig:phd} can be obtained.
Three different phases can be distinguished. In phase \two and \three
only the magnetization energy part of the current is flowing,
the current of interaction energy is zero, while the transverse
magnetization, $M^z$, is nonzero. Entering phase \one the current
of interaction energy starts to flow, while $M^z=0$ throughout
this region. On the line separating regions \one and \two ($k_h=k_\lambda$)
the symmetry of the ground state is enhanced, and it is
characterised by a single Fermi sea. There are no states above
the maximal current line, and in region \three the ground state is
the same along the $j^E=\mathrm{const.}\times h$ lines,
thus it can be represented by the $h=1$ borderline, where
the two Fermi seas merge. 
Details of the analysis of the
phase space can be found in \cite{xxwc}.
\section{Entropy of a block of spins}
\par
We are interested in the ground-state entanglement between a
block of $L$ contiguos spins and the rest of the chain.
Following Bennett \emph{et al.} \cite{Bennett} we use 
von Neumann entropy as a measure of entanglement.
It is defined as
\egy{S_L=-\mathrm{tr}(\rho_L\ln \rho_L),}
where the reduced density matrix 
$\rho_L=\mathrm{tr}_{N-L}|\Psi_g\rangle\langle \Psi_g |$ 
of the block is obtained from the ground state $|\Psi_g\rangle$
of the system by tracing out external degrees of freedom.
\par
In the calculation of the entropy we use a similar approach
that was succesfully applied in case of the XX model \cite{Kitaev}.
The first step is to introduce the fermionic operators $c_l$ and $c_l^\dag$
through the Jordan-Wigner transformation. Note, that due to the
nonsymmetric spectrum, we have to use fermionic operators instead 
of the Majorana operators that were used earlier in case of the XX model.
The ground state in our case can be completely characterised 
by the expectation values of the two-point correlations
$\langle c_m^\dag c_n \rangle=G_{mn}$; any other expectation value
can be expressed through Wick's theorem. The matrix $G$ reads
\egy{G=
\begin{bmatrix}
g_0 & g_1 & \cdots & g_{N-1}\\
g_{-1} & g_0 & & \vdots \\
\vdots & & \ddots & \vdots \\
g_{1-N} & \cdots & \cdots & g_0
\end{bmatrix},}
where the coefficients $g_l$ for an infinite chain $(N \to \infty)$
are given by
\egy{g_l=\frac{1}{2 \pi}\int_{-\pi}^{\pi}\mathrm d \theta
e^{-il\theta} \frac 1 2 \left(\frac{\Lambda_\theta}
{|\Lambda_\theta|}+1 \right),}
where $\Lambda_\theta$ is the spectrum defined in the previous
section. Note that the integrand is just the characteristic
function of the unoccupied fermionic modes.
\par
From the correlation matrix $G$ one can extract the entropy
$S_L$ of the block as follows. First, by eliminating
from $G$ the rows and columns corresponding to spins
that do not belong to the block, we obtain the correlation 
matrix $G_L$ of the state $\rho_L$:
\egy{G_L=
\begin{bmatrix}
g_0 & g_1 & \cdots & g_{L-1}\\
g_{-1} & g_0 & & \vdots \\
\vdots & & \ddots & \vdots \\
g_{1-L} & \cdots & \cdots & g_0
\end{bmatrix}.}
In principle one can reconstruct the reduced density matrix
$\rho_L$ using the matrix elements of $G_L$. However, the
entropy of the block $S_L$ can be computed in a more direct way
from the correlation matrix. Let $U \in \mathrm{SU}(L)$ denote
a unitary matrix that brings $G_L$ into a diagonal form.
This transformation defines a set of $L$ fermionic operators
$b_m=\sum_{n=1}^{L}U_{nm}c_n$ which have a diagonal
correlation matrix $\widetilde{G}_L =U^+ G_L U=\mathrm{diag}
(\lambda_1,\dots,\lambda_L)$. The expectation values are thus
\egy{\langle b_m b_n\rangle=0,\quad \langle b^\dag_m b_n\rangle=
\delta_{mn}\lambda_m,}
that is, the above fermionic modes are \emph{uncorrelated}.
Therefore the reduced density matrix can be written as
a product state
\egy{\rho_L=\rho_1 \otimes \dots \otimes \rho_L,}
where $\rho_n$ denotes the mixed state of mode $n$.
Hence the entropy $\rho_L$ is simply the sum of the
entropy of each mode:
\egy{S_L=\sum_{n=1}^L\left[-\lambda_n \ln \lambda_n-
(1-\lambda_n)\ln (1-\lambda_n)\right]\label{eq:ent}}
\subsection{Entropy asymptotics}
\begin{figure}[thb]
\includegraphics[width=0.9\columnwidth]{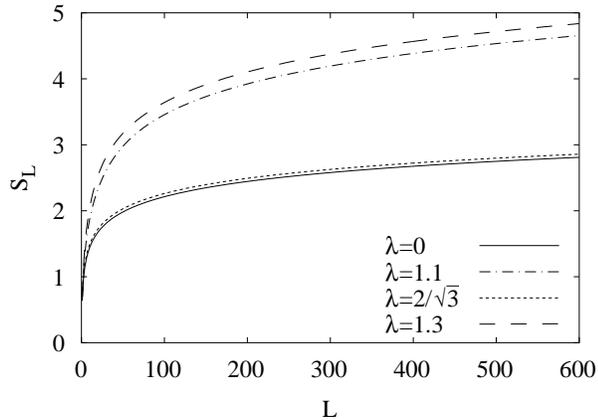}
\caption{Entropy calculated from the reduced density matrix as a 
function of the block size $L$. The magnetic field is set to $h=0.5$; the
curves correspond to different values of the current driving field 
$\lambda$. The entropy grows as $(2/3) \ln L$ in the current-carrying
phases \one ($\lambda=1.1$) and \two ($\lambda=1.3$) except
at the borderline ($\lambda=2/\sqrt 3$), where the asymptotics is
$(1/3) \ln L$, just as in the case of the XX model ($\lambda=0$).}
\label{fig:entgrowth}
\end{figure}

It follows from Eq. (\ref{eq:ent}) that in order to 
determine the entropy numerically,
one only has to diagonalize the $L \times L$ matrix $G_L$,
instead of diagonalizing the original $2^L \times 2^L$
reduced density matrix $\rho_L$. This method reduces 
considerably the computational effort, and the entropy
can be obtained for relatively large block sizes.
Fig.\ref{fig:entgrowth} shows the results of the
calculations.
\par
The ground state entropy of the XX model was first
investigated in \cite{Kitaev,Korepin} and
for $h<1$ it was found to grow asymptotically as $\frac 1 3 \ln L$
with the block size. As one starts to increase the value
of the driving field $\lambda$, the ground state (and the
entropy as well) remains the same up to the critical field $\lambda_c=1$.
Further increasing $\lambda$ one enters the current-carrying
phase \two (see Fig.\ref{fig:phd}), where the asymptotics of the
entropy changes to $\frac 2 3 \ln L$, and the same behaviour
can be observed in phase \one. The only exception is the
borderline of these phases (which is characterised by the
condition $k_h=k_\lambda$), where the entropy growth is
again $\frac 1 3 \ln L$. 
\par
For values $h \ge 1$ of the
magnetic field, all the spins are aligned in the ground state
of the XX model, thus the entropy of a block vanishes.
Switching on the current one observes
a $\frac 1 3 \ln L$ entropy asymptotics in phase \three.
\par
Summarizing the above results, one concludes that the
introduction of an energy current may lead to
a more rapid entropy growth, indicating a higher
level of ground-state entanglement. The von Neumann entropy can be
given for large block sizes ($L \to \infty$) as
\egy{S_L=\frac R 3 \ln L + S_0, \label{eq:sl}}
where $R$ is the number of Fermi seas in the spectrum
and $S_0$ is a function of the parameters $h$ and $\lambda$,
independent of $L$.
\par
The above result was obtained analytically by Keating and
Mezzadri \cite{Keating} for general quadratic
Hamiltonians that have a correlation matrix of Toeplitz type
with symmetric fermionic spectrum. 
In our case the correlation matrix is also of Toeplitz type,
but the presence of the current breaks the left-right
symmetry, resulting in a nonsymmetric spectrum.
Nevertheless, the numerical results indicate that the above
asymptotic form (\ref{eq:sl}) of the entropy seems to hold
also in this more general case.
\par
The next to leading term $S_0$ in the entropy is also given in
a closed form in \cite{Keating} for symmetric spectra.
Although the spectrum is not symmetric in our model,
there are special cases
when it can be transformed to a symmetric form.
First we note that shifting the wave numbers by $\varphi$
in the spectrum is equivalent to a unitary transformation
$V^+ G V$ of the correlation matrix, where
$V=\mathrm{diag}(1,e^{i\varphi},e^{2i\varphi},\dots,
e^{(N-1)i\varphi})$.
Since this transformation is diagonal, it leaves the
eigenvalues of the reduced density matrix $G_L$
and thus the entropy invariant.
\par
Now, if $h \lambda =1$ ($k_h+k_\lambda=\frac \pi 2$)
then the two intervals of the vacant fermionic modes
(white parts of the rectangles on
Fig.\ref{fig:phd}) have equal lengths, and a shift of 
the wave numbers by $\frac \pi 4$ symmetrizes the spectrum.
In this case the constant term $S_0$ can be expressed
as follows:
\egy{S_0= \left\{
\begin{array}{l}
\frac 2 3 \left(\ln\sqrt{4(1-\lambda^{-2})(2\lambda^{-2}-1)}+C \right),
\, k_\lambda<k_h\\
\frac 2 3 \left(\ln\sqrt{\frac{1-2\lambda^{-2}}{1-\lambda^{-2}}}+C \right),
\quad k_\lambda>k_h
\end{array},\right.\label{eq:S0}}
where $C=1+\gamma_E-6I\ln 2$ is a constant defined
through the Euler constant $\gamma_E$ and
$I\approx 0.0221603$ is a numerically evaluated integral
expression \cite{Korepin}. Thus one can see that the entropy
can be written with a scaling variable as
$S_L=\frac 2 3 (\ln \mathcal{L}+C)$, where
\egy{\mathcal{L}=\left\{
\begin{array}{l}
L \sqrt{4(1-\lambda^{-2})(2\lambda^{-2}-1)},\quad k_\lambda<k_h\\
L \sqrt{\frac{1-2\lambda^{-2}}{1-\lambda^{-2}}},\quad k_\lambda>k_h
\end{array}. \right.\label{eq:ell}}
\par
\begin{figure}[htb]
\includegraphics[width=0.7\columnwidth,angle=270]{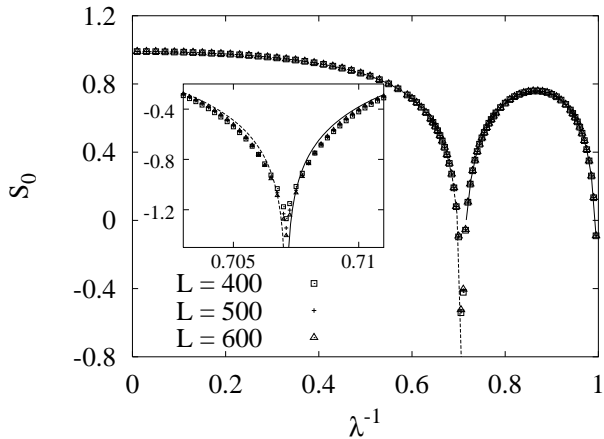}
\caption{The next to leading order part of the entropy along
the $h\lambda=1$ line for different block sizes as a function of 
$\lambda^{-1}$.
The lines are analytically calculated from the form of the spectrum.
The points perfectly fit the calculated curve except near the
vicinity of $\lambda^{-1}=1/\sqrt{2}$ (see inset) and $\lambda^{-1}=1$,
which corresponds to the boundary of the current-carrying phases.}
\label{fig:entconst}
\end{figure}

Fig.\ref{fig:entconst} shows the numerically calculated
entropy with the logarithmic part substracted for
different block sizes.
The points perfectly fit the analytically calculated
curve (\ref{eq:S0}) except near $\lambda^{-1}=\frac{1}{\sqrt{2}}$
which corresponds to the line separating the
current-carrying phases \one and \two.
Therefore, formula (\ref{eq:S0}) is applicable
when $\mathcal{L}\gg 1$.
%
\par
For nonsymmetric spectra the calculation of the
next to leading order term in the entropy is
mathematically more involved; hence we were only
able to treat the problem numerically. The
results reveal that approaching the lines characterised by
$k_h=k_\lambda$, $k_h=0$ or $k_\lambda=0$, $S_0$ seems to diverge.
This divergence is a consequence of the changing of the amplitude
of the leading term: at the boundaries of the different phases
the entropy grows as $\frac 1 3 \ln L$.
Thus approaching the boundaries, the $\frac 2 3 \ln L$
asymptotics has to be compensated with a negative
logarithmic divergence in $S_0$.
\subsection{Finite-size scaling}
\begin{figure}[htb]
\includegraphics[width=0.67\columnwidth,angle=270]{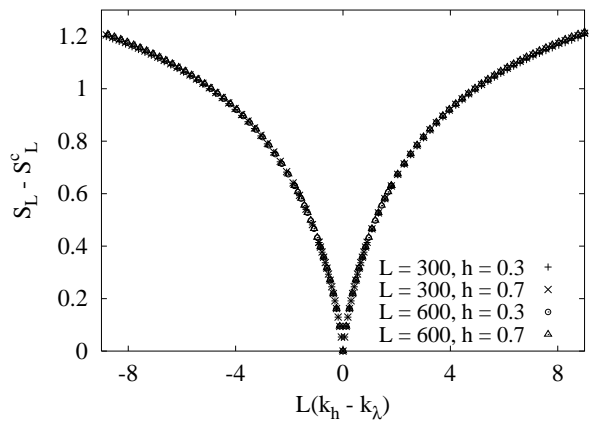}
\caption{Scaling function of the entropy near the $k_h=k_\lambda$
transition line. The ``critical'' entropy is substracted from
the total entropy and plotted against the scaling variable
$L(k_h-k_\lambda)$. The curves are calculated for different
lattice sizes ($L=300$ and $600$) and for different values of the
magnetic field ($h=0.3$ and $0.7$). The points all fit the same
scaling function.}
\label{fig:scalefunc}
\end{figure}

Obviously, for a fixed $L$, there is a finite neighbourhood
around the transition lines, where Eq.(\ref{eq:sl})
cannot be used. 
In the case of a symmetric spectrum it was seen
that it holds only when $\mathcal L \gg 1$.
If we would like to characterise the behaviour of the
entropy near these transition lines, we have to note
that we can associate a diverging length scale,
or alternatively a vanishing characteristic wave number
to each of the lines. Similarly to finite-size scaling
one writes the entropy near phase transition points
as the sum of the ``critical'' entropy and a term depending
only on the product of the block size and the characteristic
wave number. For example near the high-symmetry transition line 
($k_h=k_\lambda$) it can be written as:
\egy{S_L(k_h,k_\lambda)=S_L^c+S(L|k_h-k_\lambda|),}
where $S_L^c$ is the value of the entropy on the transition line.
The numerical calculations support the above type of scaling.
Fig.\ref{fig:scalefunc} shows the numerical results for the
scaling function near the high-symmetry line. 
Similarly, near the other transition lines 
($k_\lambda=0$ or $k_h=0$) this type of scaling is valid,
but with an other scaling function. 
%
\section{Final remarks}
We should note that analogously to the energy current,
it is possible to introduce a current of magnetization.
The resulting spectrum can be written in the same form
as that of the XX chain, however with a shift in the 
wave numbers and a decreased effective magnetic field \cite{xxwc}.
Hence, the asymptotics of the entropy will be the same;
only the $L$-independent constant increases.
Thus, interestingly, the magnetization current has a much
smaller effect on entanglement.
\par
In summary, we have shown on the example of the XX chain
that the introduction of an energy current results a more
entangled state. It would be worth considering current-carrying
steady states in other spin models, to check whether
the increase of entanglement is a general consequence
of currents.
\subsection*{Acknowledgements}
\par
We would like to thank Z. R\'acz for stimulating discussions.
This work was partially supported by OTKA Grants No. T043159,
No. T043734, and No. TS044839.

\end{document}